\documentstyle[12pt]{article}
\begin{document}
\baselineskip 0.7cm

\newcommand{\gsim}{ \mathop{}_{\textstyle \sim}^{\textstyle >} }
\newcommand{\lsim}{ \mathop{}_{\textstyle \sim}^{\textstyle <} }
\newcommand{\vev}[1]{ \left\langle {#1} \right\rangle }
\newcommand{\lsp}{ \left ( }
\newcommand{\rsp}{ \right ) }
\newcommand{\lmp}{ \left \{ }
\newcommand{\rmp}{ \right \} }
\newcommand{\llp}{ \left [ }
\newcommand{\rlp}{ \right ] }
\newcommand{\labs}{ \left | }
\newcommand{\rabs}{ \right | }
\newcommand{\KEV}{ {\rm } }
\newcommand{\MEV}{ {\rm MeV} }
\newcommand{\GEV}{ {\rm GeV} }
\newcommand{\TEV}{ {\rm TeV} }
\newcommand{\mgut}{M_{GUT}}
\newcommand{\mint}{M_{I}}
\newcommand{\mgra}{M_{3/2}}
\newcommand{\mll}{m_{\tilde{l}L}^{2}}
\newcommand{\mdr}{m_{\tilde{d}R}^{2}}
\newcommand{\mllXX}[1]{m_{\tilde{l}L , {#1}}^{2}}
\newcommand{\mdrXX}[1]{m_{\tilde{d}R , {#1}}^{2}}
\newcommand{\mgy}{m_{G1}}
\newcommand{\mgl}{m_{G2}}
\newcommand{\mgc}{m_{G3}}
\newcommand{\nuR}{\nu_{R}}
\newcommand{\slL}{\tilde{l}_{L}}
\newcommand{\slLi}{\tilde{l}_{Li}}
\newcommand{\sdR}{\tilde{d}_{R}}
\newcommand{\sdRi}{\tilde{d}_{Ri}}

\begin{titlepage}

\begin{flushright}
UT-754\\
TU-505\\
hep-ph/9607234
\\
July, 1996
\end{flushright}

\vskip 0.35cm
\begin{center}
{\large \bf Is the ee$\gamma \gamma+/\!\!\!\! E_T$ Event an Evidence of 
the Light Axino? }
\vskip 1.2cm
J.~Hisano$^{(a)}$\footnote
{Fellows of the Japan Society for the Promotion of Science.}\footnote
{The address after the next September is,
School of Physics and Astronomy, 
University of Minnesota, 
Minneapolis, MN 55455, USA.}  
, K.~Tobe$^{(a)(b)*}$,
and T.~Yanagida$^{(a)}$

\vskip 0.4cm

{\it (a) Department of Physics, University of Tokyo, Tokyo 113, Japan}
\\
{\it (b) Department of Physics, Tohoku University, Sendai 980-77,
  Japan}

\vskip 1.5cm

\abstract{
We point out that if the Peccei-Quinn symmetry breaking scale $F_a$
is in a range of the hadronic axion window ($F_a\sim 10^6$GeV), the 
ee$\gamma\gamma+/\!\!\!\! E_T$ event in the CDF experiment can be
naturally explained by a no-scale supergravity model with a light 
axino. We also stress that the hadronic axion window still survives
the intergalactic photon search, since a large entropy production due to 
the decay of Polonyi field yields a substantial dilution of the cosmic
axion density.
}

\end{center}
\end{titlepage}

%
%
%
%

The no-scale supergravity \cite{no-scale} has attracted many physicists 
in particle physics, since it may arise from a class of space-time 
compactifications in superstring theories \cite{witten}. It 
is also interesting in cosmology, since it can naturally
accommodate the chaotic inflation \cite{MSYY}, but it also provides
a consistent solution \cite{MYY} to the serious
cosmological problem in supergravity; {\it i.e.}, the Polonyi 
problem \cite{Polonyi}. However, it has been recently pointed
out \cite{KMY} that one needs a new fine tuning to solve the Polonyi problem
if the usual lightest supersymmetric particle 
(LSP) (which is perhaps a bino-dominated neutralino) is stable \cite{MYY}. 
Therefore, we are led to consider the unstable ``LSP''.
A possible way to have the unstable ``LSP'' is to break R parity. In this
case, however, we must invoke the other mechanism to avoid a rapid proton
decay. In Ref.~\cite{KMY} it has been suggested that the bino-dominated 
``LSP'' (we call it, ``bino'', hereafter) decays into the axino 
(a fermionic superpartner of the axion) \cite{RTW},
since there is a possibility that the axino remains light even after 
supersymmetry (SUSY) breaking and becomes the true LSP in the no-scale 
supergravity model
\cite{GY}.

In this letter we point out that if the Peccei-Quinn (PQ) symmetry breaking
scale $F_a$ lies in a range of the hadronic axion window ($F_a/N=(0.7-2)
\times 10^6$GeV with $N$ being the QCD anomaly factor of the PQ symmetry)
\cite{Ressell}, the ``bino'' decay into the axino can explain the 
ee$\gamma\gamma+/\!\!\!\! E_T $
event recently observed in the CDF experiment \cite{CDF}. We also stress 
that the constraint on the hadronic axion window derived from the 
intergalactic photon search \cite{MBRT}\cite{Ressell} is irrelevant in our no-scale 
supergravity model, since the decay of the Polonyi field produces a 
large amount of entropy at the late epoch of the universe evolution
and dilutes the abundance of the cosmic axion density substantially.
(The dilution factor for relativistic particles is about $10^{-13}$.)\footnote{
In Ref.~\cite{MYY} it is shown that enough baryon asymmetry is created in spite
of the large entropy production if we assume the Affleck-Dine mechanism
for baryogenesis.}

We consider, in this letter, an example of SUSY hadronic axion model 
\cite{hadronicaxion}.
The extension of our analysis based on more general models is 
straightforward. 
We assume $N$ pairs of massless new chiral superfields $\Psi_A=(Q, L)_A$ and 
$\bar{\Psi}_A=(\bar{Q}, \bar{L})_A$ ($A=1-N$) which transform as $\bf 5$ and 
$\bf 5^*$ under the grand unified gauge group 
$\rm SU(5)_{GUT}$, respectively, in addition to the SUSY standard model (SM) 
sector. 
They are assumed to have 
the PQ charge $+1$ and hence there are massless as far as the 
PQ symmetry is unbroken. All fields in the SUSY SM sector have
no PQ charge. In order to break the PQ symmetry we 
introduce a superfield $\Phi$ whose PQ charge is set as $-2$
so that the $\Phi$ can couple to the $N$ pairs of $\Psi$ and $\bar{\Psi}$
as
\begin{eqnarray}
W=\lambda_A \bar{\Psi}_A \Psi_A \Phi.
\end{eqnarray}
We assume that some physics involving the $\Phi$ field gives a nonzero
vacuum expectation value to the $\Phi$ and then 
the PQ symmetry
is spontaneously broken by the vacuum condensation $\langle\Phi \rangle
 \neq 0$.

The Nambu-Goldstone chiral multiplet arising from the PQ symmetry
breaking contains pseudo-scalar field axion $a(x)$, real-scalar field saxion 
$s(x)$, and their fermionic partner called
axino $\tilde{a}(x)$. The $N$ pairs of $\bar{\Psi}$ and $\Psi$ have masses
of $\lambda_A \langle \Phi \rangle$. The axion acquires a mass of order of 
$\Lambda_{\rm QCD}^2/\langle \Phi \rangle$ through QCD instanton effects, 
where
$\Lambda_{\rm QCD}$ is the QCD scale $\sim 100 \rm{MeV}$. On the other hand,
the axino gets a mass\footnote{
The saxion mass is given by
$m_{s}^2=  \sum_A (c/16\pi^2) \lambda_A^2 m_{SUSY}^2$ with $c \sim O(1)$.
If one takes $m_{\rm SUSY} \sim O(100) {\rm GeV}$ and $\lambda_A \sim O(0.1)$,
one gets $m_{s} \sim O(1) {\rm GeV}$. This is cosmologically harmless 
since the lifetime $\tau_{s\rightarrow gg}$ is 
$10^{-7}$ sec. for $F_a=10^6\GEV$ \cite{saxion}.
}
\begin{eqnarray}
m_{\tilde a} \simeq \sum_A\frac{1}{16 \pi^2} \lambda_A^2 m_{\rm SUSY}
\end{eqnarray}
through one loop diagrams in the no-scale supergravity model as 
shown in Ref.~\cite{MY}. Here, the $m_{\rm SUSY}$ is an induced SUSY breaking
soft mass of $\Psi$ and $\bar{\Psi}$. If one takes 
$m_{\rm SUSY} \sim O(100) {\rm GeV}$ and $\lambda_A \sim O(0.1)$, one gets 
$m_{\tilde a} \sim O(10) {\rm MeV}$. Notice that this axino $\tilde{a}$ is 
harmless 
in cosmology since the large entropy production from the Polonyi field decay
dilutes the axino density substantially. This large entropy production is also 
very important to dilute the cosmic axion density as stressed in the 
introduction. 

A crucial point in this letter is that the axion superfield 
$\Phi_a(x,\theta)$ couples to the gauge superfields through anomalies
of the PQ current as
\begin{eqnarray}
{\cal{L}}= -\sqrt{2}\frac{\alpha_i}{8 \pi} 
\int d^2 \theta \frac{\Phi_a}{F_a/N} W_\alpha^i W_\alpha^i,
\label{anomaly}
\end{eqnarray}
where $F_a = \langle \Phi \rangle$, and $W_\alpha^i$ ($i=1-3$) are
gauge superfields of the SM gauge groups U(1)$_Y$, SU(2)$_L$,
and SU(3)$_C$, and $\alpha_i$ are corresponding gauge coupling
constants ($\alpha_1= 5/3 \alpha_Y= 0.017 $, $\alpha_2 = 0.034 $, and 
$\alpha_3 = 0.12$ at the electroweak scale). This induced interactions in 
Eq.~(\ref{anomaly}) contain a bino-axino-photon coupling as
\begin{equation}
{\cal L}
 = 
- \frac{5\alpha_{em}}{24\pi} \frac{1}{\cos\theta_W} \frac{1}{F_a/N} 
\overline{\tilde{a}}\gamma_5\sigma_{\mu\nu} \tilde{B} F^{\mu\nu},
\label{interaction}
\end{equation}
from which we can estimate the decay width of 
$\tilde{B}\rightarrow \tilde{a}+ \gamma$ as
\begin{equation}
\Gamma(\tilde{B}\rightarrow \tilde{a}+ \gamma)
=\frac{25\alpha_{em}^2}{1152\pi^3}\frac{1}{\cos^2\theta_W}  \frac{M_{\tilde{B}}^3}{({F_a}/N)^2}
\end{equation}
where $M_{\tilde{B}}$ is the bino mass, and then,\footnote{
The decay $\tilde{B}\rightarrow \tilde{a}+ \gamma$ is the main decay mode.
Thus, the branching ratio ${\rm {Br}}(\tilde{B}\rightarrow \tilde{a}+ \gamma)
\simeq 100 \%$ which is also a favorable point for explaining the 
ee$\gamma \gamma+/\!\!\!\! E_T$ event.}
\begin{equation}
c\tau_{\tilde{B}} = 0.36~
                 \left(\frac{100\GEV}{M_{\tilde{B}}}\right)^3
                 \left(\frac{F_a/N}{10^6\GEV}\right)^2
                 ~{\rm c.m.}.
\end{equation}

We are now at the point of this letter. If the $F_a$ lies in the range of
hadronic axion window \cite{Ressell}, {\it i.e.}, 
\begin{equation}
F_a/N \simeq (0.7-2) \times 10^6  ~~{\rm GeV},
\end{equation}
we obtain 
\begin{equation}
c\tau_{\tilde{B}}= (0.18-1.44) ~{\rm c.m.},
\end{equation}
for $M_{\tilde{B}}=100 {\rm GeV}$.
It is now clear that the $\tilde{B}\rightarrow \tilde{a} +\gamma$ decay
can be a source of the hard photon in the ee$\gamma \gamma + /\!\!\!\!E_T$ 
event observed in the CDF experiment \cite{CDF}. 

It has been already shown in recent papers \cite{gravitino} that masses 
of a selectron $\tilde e$ and the ``bino'' $\tilde B$ must be 
$m_{\tilde e}= (80-130) {\rm GeV}$, and $M_{\tilde B}=(38-100) {\rm GeV}$ 
to explain the ee$\gamma \gamma + /\!\!\!\!E_T$ event by sequent decays;
$\tilde{e}^-( \tilde{e}^+) \rightarrow e^- (e^+)+\tilde{B}$ and $\tilde{B}
\rightarrow \gamma+ {\rm LSP}$. Let us now discuss 
the low-energy mass spectrum of SUSY particles in our model, and show that 
it is very much welcome to this event. In the no-scale supergravity model 
sfermion masses are induced by the SM gauge interactions with non-vanishing gaugino 
masses. Then, the right-handed selectron and the ``bino'' are 
naturally expected to be the lightest two among 
SUSY particles except for the axino, since they have only the U(1)$_Y$ gauge 
interaction. (See Ref.~\cite{IKYY} for a detailed calculation.)
We use renormalization group (RG) equations to evaluate a
ratio of the right-handed selectron to the bino masses. The RG equations of 
the right-handed selectron mass ($m_{\tilde{e}_R}$) and the bino mass 
($M_{\tilde{B}}$) above the mass scale of $\Psi$ and $\bar{\Psi}$ are
\begin{eqnarray}
\mu \frac{\partial m_{\tilde{e}_R}^2}{\partial\mu} &=& -8 \frac{\alpha_Y}{4\pi} M_{\tilde{B}}^2,
\nonumber\\
\mu \frac{\partial M_{\tilde{B}}}{\partial\mu} &=& 2 b_Y \frac{\alpha_Y}{4\pi} M_{\tilde{B}},
\end{eqnarray}
where $b_Y(=11+5N/3)$ is a coefficient of beta function of the U(1)$_Y$ gauge 
coupling constant and $\mu$ the renormalization point. These equations 
become those of the SUSY SM below the mass scale of $\Psi$ and $\bar{\Psi}$.
If the no-scale boundary conditions for these masses such as 
$ m_{\tilde{e}_R}=0$ is imposed at $\mu = 10^{16}$GeV, the mass ratio 
between the right-handed selectron and the ``bino'' at the electroweak scale 
is given by
\begin{eqnarray}
\frac{m_{\tilde{e}_R}}{M_{\tilde{B}}}&=& 
\left\{
\begin{array}{cc}
1.1  & (N=1) \\
1.3  & (N=2) \\
1.7  & (N=3) \\
2.6  & (N=4) 
\end{array}
\right. .
\end{eqnarray}
Here, the masses of $\Psi$ and $\bar{\Psi}$ are taken at $10^5$GeV.
The mass ratios for $N=2,3$ are suitable to the ee$\gamma 
\gamma + /\!\!\!\!E_T$ event in the CDF experiment as shown in Refs.~\cite{gravitino}\cite{BKW}\cite{DTW}.
The masses of $m_{\tilde{e}_R}$ and $M_{\tilde{B}}$ themselves take the
values around $100$ GeV to cause the correct electroweak symmetry breaking
as shown in Ref.~\cite{IKYY}, which are also desirable for the explanation
of the event. Notice that the right-handed selectron is heavier than wino 
($\tilde{W}$) for $N=4$ if the gaugino masses satisfy the GUT relation 
($M_{\tilde{W}}\simeq 2 M_{\tilde{B}}$). 
In this case, the CDF event can have another interpretation that it is a wino 
pair production accompanied with the sequent decays as 
$\tilde{W}^-(\tilde{W}^+)\rightarrow e^-(e^+)+ \bar{\nu}(\nu)+ \tilde{B}$ and 
$\tilde{W}^0 \rightarrow e^-(\nu)+ e^+ (\bar{\nu}) +\tilde{B}$, 
assuming two body decays into $W^{\pm}(Z^0) +\tilde{B}$ 
are suppressed by phase space (that is, $M_{\tilde{W}}-M_{\tilde{B}}<
m_W(m_Z)$) \cite{BKW}\cite{AKKMM}. If this interpretation is right, it is  
expected to observe multi-leptons and 2 photons events with missing 
energy.\footnote{
The decay into jets is suppressed since squarks are heavier than 
sleptons due to the SU(3)$_C$ interactions.
}
If the boundary condition is given at the gravitational scale
($\mu = 10^{18}$GeV), the mass ratio of the right-handed selectron  
to the ``bino'' becomes larger, and we can get the suitable mass spectrum
even for $N=1$ case ($m_{\tilde{e}_R}/M_{\tilde{B}}= 1.6$).

%
So far we do not have taken constraints from the effects of 
axion emission upon the life cycle of red-giant (RG) and horizontal-branch
(HB) stars in our analysis, since these constraints are based on the
statistics of small number~\cite{Ressell}. If one takes the constraints 
seriously, one obtains~\cite{Raffelt}
\begin{eqnarray}
F_a/N &>& 3 \times 10^6 ~\GEV ~~~~~(\rm{RG}),
\\
F_a/N &>& 9 \times 10^6 ~\GEV ~~~~~(\rm{HB}),
\end{eqnarray}
which is already outside the axion window. However, this problem can be 
easily solved, since the axion-photon-photon ($a \gamma \gamma$) coupling 
depends on the details of models~\cite{Choi}. For example, we assign the 
different PQ charges $Q_L$ and $Q_Q$ to the doublet $L$ and triplet $Q$,
respectively.
\footnote{Since $L$ and $Q$ have different PQ charges, the multiplets
$\Psi_A(Q,L)$ and $\bar{\Psi}_A(\bar{Q},\bar{L})$ do not form 
$\rm{SU(5)}_{\rm{GUT}}$ multiplets.}
Then, we obtain the $a \gamma \gamma$ coupling as 
\begin{eqnarray}
{\cal{L}}_{a \gamma \gamma}&=&\frac{\kappa}{4} 
a F_{\mu \nu} \tilde{F}^{\mu \nu},
\\
\kappa&=&\frac{\alpha_{em}}{2\pi}\frac{1}{F_a/N}
\left[ \frac{2}{3}(1+3 \gamma)-\frac{2(4+z)}{3(1+z)}\right],
\label{agammagamma}
\end{eqnarray}
where $\gamma$ is the ratio of the PQ charge, $\gamma=Q_L/Q_Q$ and $z$ the
mass ratio of up- and down-quarks, $z=m_u/m_d$. Notice that the second 
term in the bracket of Eq.(\ref{agammagamma}) denotes the contribution 
from the long-distance effect.
We can see that $a \gamma \gamma$ coupling is almost vanishing for 
$z \simeq 0.56$ and $\gamma \simeq 2/3$. In this case, the above constraints
from RG and HB stars become weaker~\cite{Ressell}
\begin{eqnarray}
F_a/N  &>& 0.2 \times 10^6 ~\GEV ~~~~~(\rm{RG}),
\\
F_a/N  &>& 0.6 \times 10^6 ~\GEV ~~~~~(\rm{HB}),
\end{eqnarray}
since subject to the constraints is the $a \gamma \gamma$ coupling 
strength but not $F_a$ itself. On the other hand, the bino-axino-photon 
($\tilde{B}\tilde{a} \gamma$) coupling does not have
the contribution from the long-distance effect and hence there is not such a
cancellation in the $\tilde{B}\tilde{a} \gamma$ coupling. Thus, the analysis
in this paper is unchanged.

In summary, we argue that the hadronic axion window is not yet
excluded by the intergalactic photon search, since there is
a substantial dilution of the cosmic axion density in the no-scale 
supergravity model. 
If the PQ symmetry breaking scale is in the hadronic axion window,
the ee$\gamma\gamma+ /\!\!\!\! E_T$ event in the CDF experiment can be 
naturally interpreted 
as a result of the cascade decays; $\tilde{e}_R^-(\tilde{e}_R^+)\rightarrow
e^-(e^+) + \tilde{B}$ and $\tilde{B}\rightarrow {\rm axino} + \gamma$.
We hope that this hadronic axion window will be tested 
by future axion searches \cite{moriyama}.

\vskip 1.5cm
\noindent{\bf Acknowledgment}

We thank K.~Choi for informing the presence of Ref.~\cite{Choi}.

\newpage

%
%
\newcommand{\Journal}[4]{{\sl #1} {\bf #2} {(#3)} {#4}}
\newcommand{\PL}{\sl Phys. Lett.}
\newcommand{\PR}{\sl Phys. Rev.}
\newcommand{\PRL}{\sl Phys. Rev. Lett.}
\newcommand{\NP}{\sl Nucl. Phys.}
\newcommand{\ZP}{\sl Z. Phys.}
\newcommand{\PTP}{\sl Prog. Theor. Phys.}
\newcommand{\NC}{\sl Nuovo Cimento}

%

\end{document}